\documentclass[preprint2]{aastex631}
\usepackage{xcolor}
\usepackage{CJKutf8}
\usepackage[normalem]{ulem}

\begin{document}
\title{Enhanced Merger Fractions in a Reionization-Era Protocluster}
\author{Lorraine C. Marcelin}
\affiliation{Steward Observatory, University of Arizona, 933 N. Cherry Ave, Tucson, AZ 85721, USA}
\author[0000-0002-6184-9097]{Jaclyn~B.~Champagne}
\affiliation{Steward Observatory, University of Arizona, 933 N. Cherry Ave, Tucson, AZ 85721, USA}
\author[0000-0002-7633-431X]{Feige Wang}
\affiliation{Steward Observatory, University of Arizona, 933 N. Cherry Ave, Tucson, AZ 85721, USA}
\affiliation{Department of Astronomy, University of Michigan, 1085 S. University Ave., Ann Arbor, MI 48109, USA}
\author[0000-0003-3310-0131]{Xiaohui Fan}
\affiliation{Steward Observatory, University of Arizona, 933 N. Cherry Ave, Tucson, AZ 85721, USA}
\author[0000-0003-4924-5941]{Maria Pudoka}\affiliation{Steward Observatory, University of Arizona, 933 N. Cherry Ave, Tucson, AZ 85721, USA}
\author[0000-0003-0747-1780]{Wei Leong Tee}
\affiliation{Steward Observatory, University of Arizona, 933 N. Cherry Ave, Tucson, AZ 85721, USA}
\author[0000-0003-3307-7525]{Yongda Zhu}
\affiliation{Steward Observatory, University of Arizona, 933 N. Cherry Ave, Tucson, AZ 85721, USA}

\begin{abstract}

Mergers play a critical role in galaxy evolution, but their relationship to their surrounding environments is unexplored at high redshift.  We investigate the galaxy merger rate among 124 [OIII] emitters at $5.3<z<6.9$ as a function of local galaxy density.  Identified in the ASPIRE JWST/NIRCam grism survey,  we investigate three density regimes: a $z=6.6$ quasar-centered protocluster, two overdensities at $z=5.4$ and $z=6.2$, and field galaxies.
We evaluate merger candidates through close pair and morphological criteria in NIRCam imaging, finding that the $z=6.6$ protocluster contains the highest fraction of galaxies meeting either criterion.
We observe a $>3\sigma$ enhancement of the merger fraction amongst all three overdense structures compared to the field.
Eleven galaxies are classified as ``active mergers" satisfying both merger criteria, 
all of which occur within the overdensity samples.
We conclude that environment affects the merger rates of galaxies at $z>6$, leading to increased specific star formation at the 4$\sigma$ level.
\end{abstract}

\section{Introduction} \label{sec:intro}

Mergers can alter a galaxy's structural evolution and star formation rate.
JWST/NIRCam's high spatial resolution in imaging, wide-field slitless spectroscopy and access to the rest optical at high redshift, enable us to observe the UV/optical morphology and provide precise redshifts of galaxies at $z>6$, allowing for detailed studies of mergers.
The JWST Cycle 1 ASPIRE survey \citep{Wang} obtained F356W NIRCam/WFSS observations of 25 $z>6.5$ quasar fields to investigate their host galaxies and environments.
\citet{Champagne2024a} and \citet{Wang} identified a protocluster containing 53 H$\beta$+[OIII] emitters extending beyond 10 Mpc centered on a luminous $z=6.6$ quasar, J0305$-$3150. 
The sample contains two serendipitous line-of-sight overdensities, Overdensity 1 at $z=5.35-5.41$ (19 galaxies spanning 10 transverse Mpc), and Overdensity 2 at $z=6.2-6.3$ (18 galaxies spanning 16 Mpc on sky).
The total sample contains 124 galaxies, including 34 field galaxies at $5.3<z<6.9$.
 
Simulations predict that brightest cluster galaxies form and evolve through frequent mergers at $z>4$ \citep{Rennehan}, while observations show that global merger rate and star formation peak in cosmic noon ($2<z<4$) protoclusters \citep{Hine}.
However, to date, we have not been able to determine if merger rates are already enhanced in reionization-era protoclusters.
In this note, we identify potential merger candidates in a $z>6$ sample to determine if mergers occur more frequently in denser environments compared to the field. 
We use a $\Lambda$CDM cosmology with $\Omega_{\Lambda}$=0.7, $\Omega_M$=0.3, and $H_0$=70 km\,s$^{-1}$\,Mpc$^{-1}$.

\section{Analysis} \label{Analysis}
\
\
We identify potential mergers among the [OIII] emitters through two criteria: the close pair rate and a morphological classification.

\subsection{Non-Parametric Morphological Parameters\label{subsec:Non Parametric Data}}

To evaluate the morphological parameters of our galaxies, we constructed 101$\times$101 pixel cutouts of the F356W images and corresponding segmentation maps at the center coordinates of each galaxy, tracing H$\beta$+[OIII]  emission. 
We use \verb|source_morphology| contained in the \verb|statmorph| Python package, which takes the image cutout, segmentation map, and PSF as input and returns non-parametric morphological parameters including the circularized half-radius, asymmetry, concentration, Gini, and M$_{20}$ \citep{Rodriguez-Gomez}. 
We used the segmap to identify and mask out potential contaminants within the cutout. 
We discarded 7 galaxies whose signal to noise ratio $S/N<2.5$  were deemed unreliable by \verb|statmorph| for morphological measurements \citep{Lotz}. 

The Gini coefficient measures the imbalance in light distribution across the galaxy, while $M_{20}$ determines the brightest pixels in the galaxy as a function of galactocentric distance. Thus having both high Gini and $M_{20}$ values indicates that the galaxy is a potential merger. We classify mergers following \citet{Costantin}:
\begin{equation}
    G >  -0.14 \times M_{20}\ + 0.33
    \label{eq3}
\end{equation} 

\subsection{Identifying Close Pairs\label{subsec: Identifying Merger Signatures}}

Using the close pair criterion from \citet{Ventou}, Equation \ref{eq1}  
takes into account the separation distance $r_{p}$,  the angular separation 
$\theta$; 
the angular diameter distance $d_a$; and the mean redshift $z_m$ between two galaxies. Equation \ref{eq2} considers the difference in rest frame velocity where $c$ is the speed of light. Galaxies qualify as mergers if the pair separation $r_{p}^{max}\leqslant 50$\,kpc and $\Delta v \leqslant 315$\, km\,s$^{-1}$  or $r_{p}^{max}\leqslant 100$\, kpc and $\Delta$v$\leqslant100$ \,km\,s$^{-1}$:

\begin{equation}  \theta \times d_a(z_m) \leqslant r_{p}^{max} 
\label{eq1}
\end{equation}
\
\begin{equation}
\Delta v = c \times \frac{|z_{1}-z_{2}|}{z_{m}}
\label{eq2}
\end{equation}

\begin{figure*}[h!]
    \centering
    \gridline{
    \fig{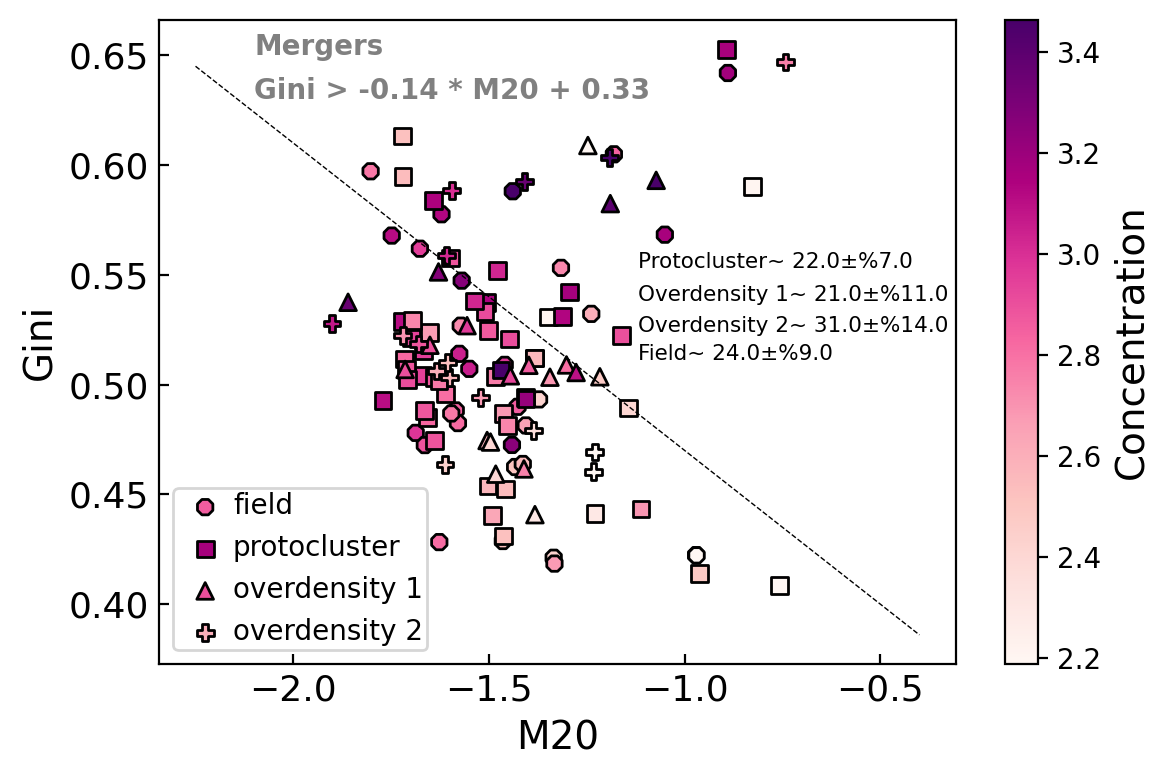}{0.5\textwidth}{(a)}
    \fig{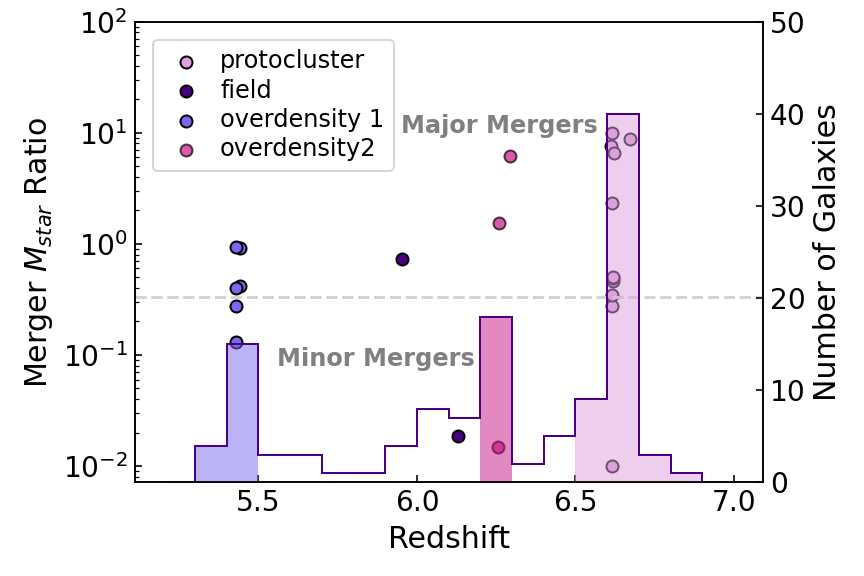}{0.5\textwidth}{(b)}
    }
    \gridline{
    \fig{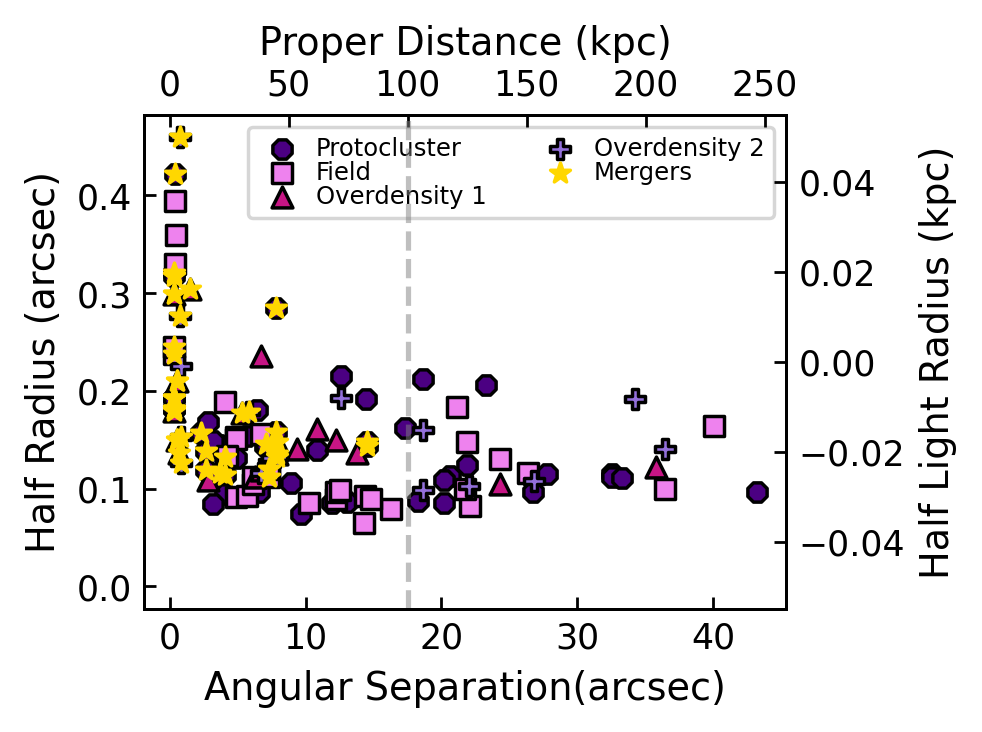}{0.52\textwidth}{(c)}
\fig{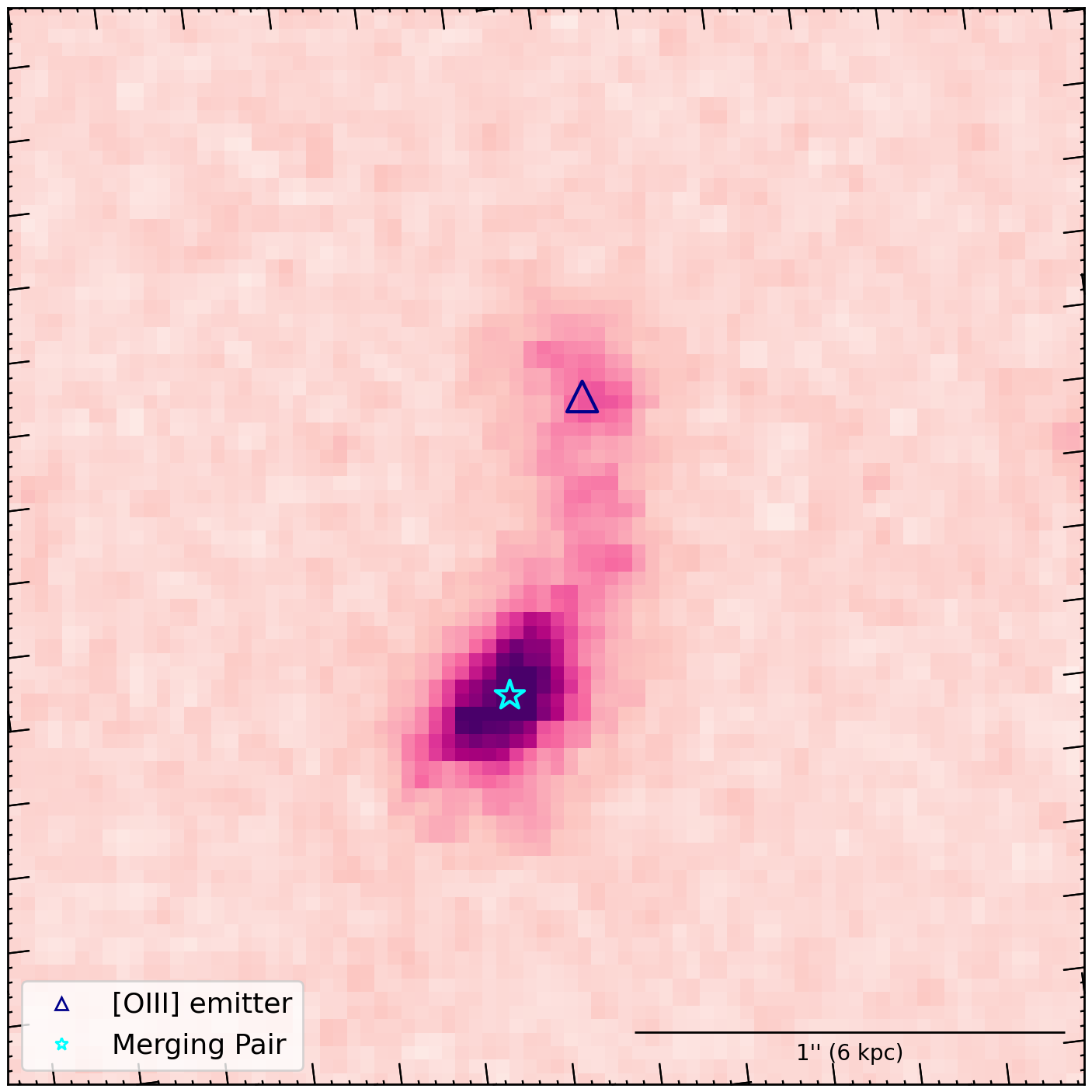}{0.35\textwidth}{(d)}
}

    \caption{\textit{a.} Gini vs. $M_{20}$ in each subsample, 
    color coded by concentration. Above the dotted line, galaxies are consistent with a merger scenario \citep{Costantin}. The merger fractions for each subsample are listed. 
    \textit{b.} The stellar mass ratio of each merger pair (left axis) plotted against median redshift. The redshift histogram of the full sample is overlaid (right axis), showing which mergers belong to overdensities (shaded) or field (empty). The dotted line shows the 1:3 mass ratio between major and minor mergers.
    \textit{c.} The half-light radius of the galaxies versus angular distance to its nearest neighbor. 
    Yellow stars signify the close-pair mergers. The dashed line is the minimum  distance criterion at $ r_{p}^{max}\geqslant 100$\,kpc.
    {$d.$} A 2\arcsec$\times$2\arcsec\, F356W cutout showing two markers at the center coordinates of a galaxy and its merging pair. The [OIII] emitter pair satisfies both dynamical and morphological criteria. 
    }
    \label{fig:1}
\end{figure*}

\section{Results} \label{Results}
Of the sample of 117 galaxies, 44 met the close pair rate qualifications with a total merger fraction of 38$\pm$7\%,  assuming Poisson errors. 
22 were in the $z=6.6$ protocluster, 12 were in Overdensity 1 at $z=5.4$, 6 in Overdensity 2 at $z=6.1$, and 4 were in the field, yielding specific merger rates of 45$\pm$10\%, 63$\pm$10\%, 38$\pm$20\%, and 12$\pm$6\% respectively.
This shows clear evidence of an enhancement of close pairs within overdensities.

Figure \ref{fig:1}a shows the Gini-$M_{20}$ relationship between individual galaxies. 
Galaxies in the protocluster had the highest Gini and $M_{20}$ values, consistent with most mergers residing in the protocluster. From the morphological criteria, 28/117 galaxies were mergers --- a total fraction of 24$\pm$5\%. 
Of the Gini-M$_{20}$ merger candidates 11 were in the protocluster, 4 in Overdensity 1, 5 in Overdensity 2, and 8 in the field. The discrepancies between the two may indicate a diversity in the merger stages to which the criteria are sensitive, though 
the merger fractions are consistent within $2\sigma$. We classified  11 galaxies as ``active mergers" having satisfied both the structural and close pair criteria, all of which were in overdense regions. 

Next we calculated the stellar mass ratio of the mergers \citep[derived through SED fitting by][]{Champagne2024b}  in each overdensity group to determine the frequency of minor versus major mergers (defined as $M_*$ ratios greater than 1:3) that occur in any of the overdensity samples. 
Figure \ref{fig:1}b shows that the mass ratio and number of major mergers increased with redshift in all overdensities. 
The median specific star formation rate for non-mergers is 0.32 $\pm$ 0.03 Gyr$^{-1}$\, and 0.65$\pm$0.05 Gyr$^{-1}$\, for mergers,
suggesting that the merging process 
leads to increased star formation. 
The relationship between the half-radius and the angular distance between pairs is shown in Figure \ref{fig:1}c, where the half-light radius of galaxies decreases with separation. 
Most galaxy pairs with very close separations are larger, which could indicate early stages of morphological disturbances due to mergers. 

\section{Discussion \& Conclusion} \label{Conclusion} 
In this work, we analyzed the relationship between galaxy merger rate and environmental density, classifying them through both morphological and close-pair criteria.
We found a $\sim3\sigma$ enhancement of mergers in the protocluster, with marginal evidence for an enhancement in the other two overdensities compared to the field. 
Evidence of morphological disturbance is present in many, but not all, of the close pair merger candidates. This suggests that the separation criteria is more sensitive to galaxies being on track to merge in the near future given their small velocity  separations, but remain too far apart for the morphology to be affected gravitationally yet. 
The \citet{Ventou} criteria are capable of identifying future and early-stage mergers. Those that only satisfied the Gini-$M_{20}$  correlation could be ongoing or post-mergers that no longer meet the velocity requirements as separate galaxies. . 
Galaxies satisfying bothcriteria are therefore at a midpoint of the merging process, and these were only identified in the overdensities. One of the ``active mergers" is shown in panel $d$ of Figure \ref {fig:1} where the irregular shape and proximity to its merging pair is evident. 
The compactness of the overdensity also correlates with the merger fraction, as the sparser protocluster and Overdensity 2 had lower merger fractions than Overdensity 1.

The stellar mass ratios suggest major mergers are more common in denser environments, particularly evident in the protocluster. The merging process increases the specific star formation rate in these galaxies, consistent with simulation predictions that protoclusters evolve through frequent mergers at $z>4$ \citep{Rennehan}.
Evidence of significantly enhanced merger fractions in more mature $z\sim2.5$ protoclusters was observed by \citet{Giddings}, nearly 2 Gyr after the structures discussed here. Here we present the first evidence that close pair rates are enhanced even at $z>6$, despite the smaller density contrast between protocluster and field at this epoch.

Overall, the increased merger rate in the Protocluster, Overdensity 1 and Overdensity 2 reveals an enhanced merger fraction in denser environments during the nascent stage of protocluster development.
Combined with the specific star formation rate of galaxies being higher for merging galaxies, we conclude that densely populated environments correlate with increased merger rates and therefore enhanced star formation rates, representing accelerated evolution of galaxies as a function of density.

\section*{Acknowledgments}
JBC acknowledges funding from the JWST Arizona/Steward Postdoc in Early galaxies and Reionization (JASPER) Scholar contract at the University of Arizona. FW and JBC acknowledge support from NSF Grant AST-2308258. This work is based on observations made with the NASA/ESA/CSA James Webb Space Telescope. The data were obtained from the Mikulski Archive for Space Telescopes at the Space Telescope Science Institute, which is operated by the Association of Universities for Research in Astronomy, Inc., under NASA contract NAS
5-03127 for JWST. These observations are associated with programs \#2078 and \#3225. Support for these programs was given through a grant from the Space Telescope Science Institute, which is operated by the Association of Universities for Research in Astronomy, Inc., under NASA contract NAS 5-03127.

\bibliography{main}

\begin{thebibliography}{}
\expandafter\ifx\csname natexlab\endcsname\relax\def\natexlab#1{#1}\fi
\providecommand{\url}[1]{\href{#1}{#1}}
\providecommand{\dodoi}[1]{doi:~\href{http://doi.org/#1}{\nolinkurl{#1}}}
\providecommand{\doeprint}[1]{\href{http://ascl.net/#1}{\nolinkurl{http://ascl.net/#1}}}
\providecommand{\doarXiv}[1]{\href{https://arxiv.org/abs/#1}{\nolinkurl{https://arxiv.org/abs/#1}}}

\bibitem[{{Champagne} {et~al.}(2025{\natexlab{a}}){Champagne}, {Wang}, {Zhang}, {Yang}, {Fan}, {Hennawi}, {Sun}, {Ba{\~n}ados}, {Bosman}, {Costa}, {Eilers}, {Endsley}, {Jin}, {Jun}, {Li}, {Lin}, {Liu}, {Loiacono}, {Lupi}, {Mazzucchelli}, {Pudoka}, {Protu{\v{s}}ov{\`a}}, {Rojas-Ruiz}, {Tee}, {Trebitsch}, {Venemans}, {Zhuang}, \& {Zou}}]{Champagne2024a}
{Champagne}, J.~B., {Wang}, F., {Zhang}, H., {et~al.} 2025{\natexlab{a}}, \apj, 981, 113, \dodoi{10.3847/1538-4357/adb1bd}

\bibitem[{{Champagne} {et~al.}(2025{\natexlab{b}}){Champagne}, {Wang}, {Yang}, {Fan}, {Hennawi}, {Sun}, {Ba{\~n}ados}, {Bosman}, {Costa}, {Habouzit}, {Jin}, {Jun}, {Li}, {Liu}, {Loiacono}, {Lupi}, {Mazzucchelli}, {Pudoka}, {Rojas-Ruiz}, {Tee}, {Trebitsch}, {Zhang}, {Zhuang}, \& {Zou}}]{Champagne2024b}
{Champagne}, J.~B., {Wang}, F., {Yang}, J., {et~al.} 2025{\natexlab{b}}, \apj, 981, 114, \dodoi{10.3847/1538-4357/adb1bc}

\bibitem[{{Costantin} {et~al.}(2024){Costantin}, {Gillman}, {Boogaard}, {P{\'e}rez-Gonz{\'a}lez}, {Iani}, {Rinaldi}, {Melinder}, {Crespo G{\'o}mez}, {Colina}, {Greve}, {{\"O}stlin}, {Wright}, {Alonso-Herrero}, {{\'A}lvarez-M{\'a}rquez}, {Annunziatella}, {Bik.}, {Caputi}, {Dicken}, {Eckart}, {Hjorth}, {Ilbert}, {Jermann}, {Labiano}, {Langeroodi}, {Pei{\ss}ker}, {Pye}, {Tikkanen}, {van der Werf}, {Walter}, {Ward}, {G{\"u}del}, \& {Henning}}]{Costantin}
{Costantin}, L., {Gillman}, S., {Boogaard}, L.~A., {et~al.} 2024, arXiv e-prints, arXiv:2407.00153, \dodoi{10.48550/arXiv.2407.00153}

\bibitem[{{Giddings} {et~al.}(2025){Giddings}, {Lemaux}, {Forrest}, {Shen}, {Sikorski}, {Gal}, {Cucciati}, {Golden-Marx}, {Ronayne}, {Shah}, {Amor{\'\i}n}, {Bardelli}, {Baxter}, {Cassar{\`a}}, {De Lucia}, {Fontanot}, {Gururajan}, {Hathi}, {Hirschmann}, {Hung}, {Lubin}, {Sanders}, {Vergani}, {Xie}, \& {Zucca}}]{Giddings}
{Giddings}, F., {Lemaux}, B.~C., {Forrest}, B., {et~al.} 2025, arXiv e-prints, arXiv:2503.04913, \dodoi{10.48550/arXiv.2503.04913}

\bibitem[{{Hine} {et~al.}(2016){Hine}, {Geach}, {Alexander}, {Lehmer}, {Chapman}, \& {Matsuda}}]{Hine}
{Hine}, N.~K., {Geach}, J.~E., {Alexander}, D.~M., {et~al.} 2016, \mnras, 455, 2363, \dodoi{10.1093/mnras/stv2448}

\bibitem[{{Lotz} {et~al.}(2006){Lotz}, {Madau}, {Giavalisco}, {Primack}, \& {Ferguson}}]{Lotz}
{Lotz}, J.~M., {Madau}, P., {Giavalisco}, M., {Primack}, J., \& {Ferguson}, H.~C. 2006, \apj, 636, 592, \dodoi{10.1086/497950}

\bibitem[{{Rennehan} {et~al.}(2020){Rennehan}, {Babul}, {Hayward}, {Bottrell}, {Hani}, \& {Chapman}}]{Rennehan}
{Rennehan}, D., {Babul}, A., {Hayward}, C.~C., {et~al.} 2020, \mnras, 493, 4607, \dodoi{10.1093/mnras/staa541}

\bibitem[{{Rodriguez-Gomez} {et~al.}(2019){Rodriguez-Gomez}, {Snyder}, {Lotz}, {Nelson}, {Pillepich}, {Springel}, {Genel}, {Weinberger}, {Tacchella}, {Pakmor}, {Torrey}, {Marinacci}, {Vogelsberger}, {Hernquist}, \& {Thilker}}]{Rodriguez-Gomez}
{Rodriguez-Gomez}, V., {Snyder}, G.~F., {Lotz}, J.~M., {et~al.} 2019, \mnras, 483, 4140, \dodoi{10.1093/mnras/sty3345}

\bibitem[{{Ventou} {et~al.}(2019){Ventou}, {Contini}, {Bouch{\'e}}, {Epinat}, {Brinchmann}, {Inami}, {Richard}, {Schroetter}, {Soucail}, {Steinmetz}, \& {Weilbacher}}]{Ventou}
{Ventou}, E., {Contini}, T., {Bouch{\'e}}, N., {et~al.} 2019, \aap, 631, A87, \dodoi{10.1051/0004-6361/201935597}

\bibitem[{{Wang} {et~al.}(2023){Wang}, {Yang}, {Hennawi}, {Fan}, {Sun}, {Champagne}, {Costa}, {Habouzit}, {Endsley}, {Li}, {Lin}, {Meyer}, {Schindler}, {Wu}, {Ba{\~n}ados}, {Barth}, {Bhowmick}, {Bieri}, {Blecha}, {Bosman}, {Cai}, {Colina}, {Connor}, {Davies}, {Decarli}, {De Rosa}, {Drake}, {Egami}, {Eilers}, {Evans}, {Farina}, {Haiman}, {Jiang}, {Jin}, {Jun}, {Kakiichi}, {Khusanova}, {Kulkarni}, {Li}, {Liu}, {Loiacono}, {Lupi}, {Mazzucchelli}, {Onoue}, {Pudoka}, {Rojas-Ruiz}, {Shen}, {Strauss}, {Tee}, {Trakhtenbrot}, {Trebitsch}, {Venemans}, {Volonteri}, {Walter}, {Xie}, {Yue}, {Zhang}, {Zhang}, \& {Zou}}]{Wang}
{Wang}, F., {Yang}, J., {Hennawi}, J.~F., {et~al.} 2023, \apjl, 951, L4, \dodoi{10.3847/2041-8213/accd6f}

\end{thebibliography}
\end{document}